\documentclass[amsmath,aip,reprint,superscriptaddress,longbibliography]{revtex4-1}

\usepackage{bbold}
\usepackage{graphics}
\usepackage{graphicx}
\usepackage{amsthm}
\usepackage{amsmath}
\usepackage{wasysym}
\usepackage{amssymb}
\usepackage{xfrac}
\usepackage{color}
\usepackage[dvips]{epsfig}
\usepackage{subfigure}
\newcommand{\bs}{\boldsymbol}

\begin{document}

\title{Topological Phases in Oxide Heterostructures with Light and Heavy Transition Metal Ions}

\author{Gregory A. Fiete}
\email{fiete@physics.utexas.edu}%
\affiliation{Department of Physics, The University of Texas at Austin, Austin, Texas 78712, USA}
\author{Andreas R\"uegg}
\affiliation{Theoretische Physik, ETH Z\"urich, CH-8093 Z\"urich, Switzerland}
\date{\today}

\begin{abstract}
Using a combination of density functional theory, tight-binding models, and Hartree-Fock theory, we predict topological phases with and without time-reversal symmetry breaking in oxide heterostructures.  We consider both heterostructures containing light transition metal ions, and those containing heavy transition metal ions.  We find the (111) growth direction naturally leads to favorable conditions for topological phases in both perovskite structures and pyrochlore structures.  For the case of light transition metal elements, Hartree-Fock theory predicts the spin-orbit coupling is effectively enhanced by on-site multiple-orbital interactions and may drive the system through a topological phase transition, while heavy elements with intrinsically large spin-orbit coupling require much weaker, or even vanishing electron interactions to bring about a topological phase.  
\end{abstract}

\maketitle

\section{Introduction}
Topological insulators (TI) have now been studied for nearly a decade and there are a number of excellent reviews available, from both an experimental and theoretical perspective.\cite{Moore:nat10,Hasan:rmp10,Qi:rmp11,Ando:jpsj13}  Aside from the important example of SmB$_6$,\cite{dzero2013,Dzero:prl10,Neupane:nc13,Zhang:prx13} 
all other experimental examples of topological insulators are weakly correlated.  The most important reason for the weak correlation effects is that the bands of topological insulators near the Fermi energy are derived from $s$ and $p$-type orbitals, which are rather extended.  By contrast, the bands about the Fermi energy in SmB$_6$ are predominantly derived from $f$-orbitals, which are more localized, and therefore lead to flatter bands and enhanced interaction effects.\cite{Lu:prl13}  

One of the persistent challenges in experimental studies of TI  is the problem of high bulk conductivity.\cite{Ando:jpsj13}  While some progress has been made on this front over the past few years, it has largely been incremental, and for the most part has been focused on bismuth-based TI.  An alternative route is to look for new classes of strongly insulating materials that might support topological insulator phases.  If one is also interested in studying interaction effects that could possibly take one beyond the ``band theory" description of topological insulators, materials with orbitals more localized than the $s$ and $p$-orbitals are highly desirable.  Transition metal oxides, which typically have bands derived from $d$-orbitals close to the Fermi energy, are excellent candidates:  There are a large number of insulating oxides, and interaction effects are known to be important in many of them--the high temperature cuprate superconductors serving as an excellent example.\cite{Imada:rmp98,Lee:rmp06}  Indeed, there have been a number of theoretical proposals for strongly correlated topological insulators in transition metal oxide systems.\cite{Pesin:np10,Kargarian:prb11,Kargarian:prl13,Witczak-Krempa:prb10,Maciejko:prb13,Ruegg:prl12,Maciejko:prl14,Go:prl12,Shitade:prl09}  A recent review of the prospects for such exotic phases in the context of three dimensional iridium (and other heavy transition metals) oxides is given in Ref.[\onlinecite{witczak-krempa2014}].

In this article, we focus on topological phases in thin film (two-dimensional) oxide heterostructures that can be described within the band theory picture.  There are three prime candidates: (1) The time-reversal invariant topological insulator characterized by a single $Z_2$ invariant, (2) The time-reversal symmetry broken Chern insulator characterized by a quantized Chern number and quantized Hall conductance, and (3) The mirror symmetry protected (with respect to the center of the plane of the thin film\cite{Liu:nm14}) topological crystalline insulator characterized by a mirror Chern number.\cite{Teo:prb08,Hsieh:naC2012,Liu_TCI:prb13,Hsieh:prb14}  Combinations of these are possible as well for a ``doubly topological" system, though clearly (1) and (2) are mutually exclusive. The Chern insulator differs from the quantum Hall insulator in that the former has time-reversal symmetry spontaneously broken by interactions (that is, a spontaneous magnetization of some sort), while the latter has time-reversal symmetry broken by the application of an external magnetic field.  As a result, a Chern insulator requires interactions (because there is no spontaneous magnetism without electron-electron interactions), while a (integer) quantum Hall system does not.  A topological crystalline insulator may possess time-reversal symmetry or have it broken by magnetic order; the only restriction is that any magnetic order present must respect the mirror symmetry.\cite{Hsieh:naC2012}

In the remainder of this article, we will consider two transition metal oxide structures, the perovskite with formula ABO$_3$ and the pyrochlore oxide A$_2$B$_2$O$_7$, where A is typically a rare-earth element, B is a transition metal element, and O is oxygen.  Two specific examples we study are LaNiO$_3$ and Y$_2$Ir$_2$O$_7$, which are both materials that have been grown and are well characterized in bulk form.  Our new angle is to study the properties of thin films of these materials that are grown along the (111) crystalline axis.  We find the physical properties of these films are rather different from the bulk.  Moreover, there does not appear to be an obvious way to infer the film properties from the bulk.  As a result, these systems appear to be excellent candidates for exploring novel phenomena, such as topological phases, even when the bulk (non-thin film) materials possess strikingly different properties, such as a conducting behavior.  We are thus presented with the exciting possibility of finding ``new physics" in ``old materials".

\section{Thin Film Oxide Heterostructures}

\begin{figure}[h]
\includegraphics[width=1\linewidth]{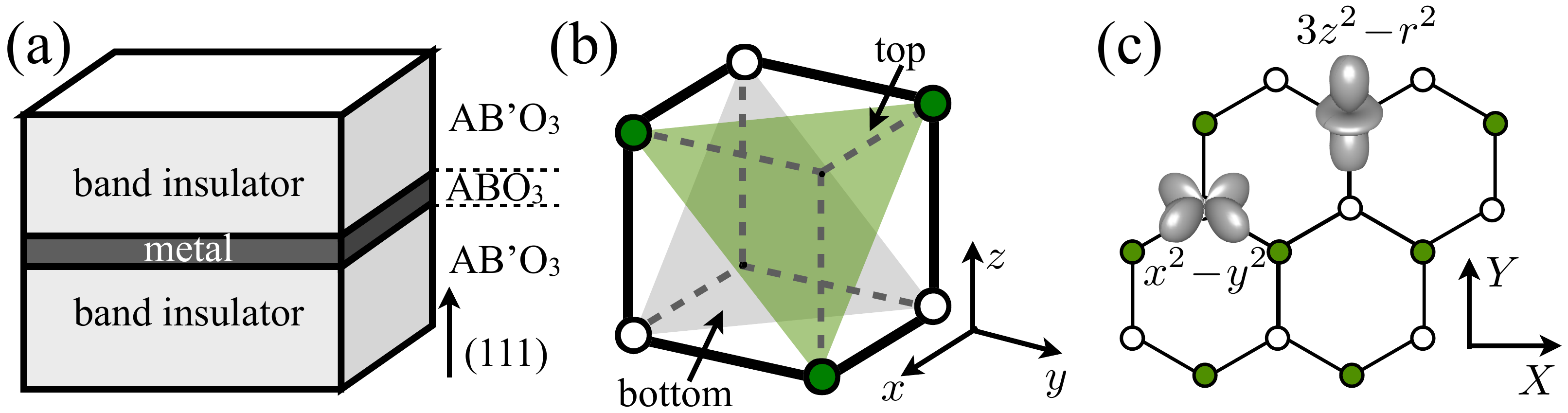}
\caption{(Color online.) (a) We consider oxide heterostructures grown along the (111) direction of the form AB'O$_3$/ABO$_3$/AB'O$_3$ for the perovskite structure and A$_2$B'$_2$O$_7$/A$_2$B$_2$O$_7$/A$_2$B'$_2$O$_7$ for the pyrochlore structure.  Both AB'O$_3$ and A$_2$B'$_2$O$_7$ are assumed to be non-magnetic, wide band gap (band) insulators. The shaded region consists of a (111) bilayer of the metallic ABO$_3$ perovskite, such as LaNiO$_3$, or in the case of pyrochlore structures, a bilayer or a trilayer of a material such as Y$_2$Ir$_2$O$_7$.  (b) Shown are the locations of the transition-metal ions (B) for the ABO$_3$ structure. (c) The ABO$_3$ bilayer system forms a ``buckled honeycomb" lattice, which appears as a honeycomb lattice when projected to the plane perpendicular to (111). We assume that the relevant orbital degrees of freedom are the $e_g$ orbitals of the transition-metal ions for ABO$_3$.  For A$_2$B$_2$O$_7$ with heavy transition metal ions, the relevant orbitals come from the spin-orbit split $t_{2g}$ manifold in the class of materials we study in this paper. }
\label{fig:fig1}
\end{figure}

We are interested in a heterostructure similar to that shown in Fig.\ref{fig:fig1}, where the film is grown along the (111) direction.  Most experimental studies of thin films in the perovskite systems ABO$_3$ are grown along the (001) (or symmetry related) direction because aligning the growth direction along a cubic axis generally favors high quality films.  On the other hand, growing along (111) effectively changes the crystal structure in a single layer thin film from a square lattice of transition metal ions for (001) to a triangular lattice of transition metal ions for (111), as shown in Fig.\ref{fig:fig1}.  For a bilayer grown along (001), one will have two square lattices of transition metal ions stacked directly on top of each other, but for growth along (111) a second ``shifted" triangular lattice will sit on top of the original one.  The combination of the two shifted triangular lattices is a ``buckled" honeycomb lattice.  In this example, growing along the (111) direction allows one to effectively ``engineer" the lattice of the thin film, with important implications for the band structure in the weak coupling limit and the magnetic order in the strong coupling limit.

Likewise, the (111) growth direction for the A$_2$B$_2$O$_7$ pyrochlores leads to alternating planes of triangular and kagome lattices for the transition metal ions. [See Fig.\ref{fig:pyrochlore}(a).]  To the best of our knowledge, there have been no published experimental results on bilayer or trilayer  films of A$_2$B$_2$O$_7$ grown along (111), though there are a number of systems where (111) growth of ABO$_3$ films has been demonstrated, \cite{Middey:apl12,Blok:apl11,Middey14} and also of AB$_2$O$_4$ spinels.\cite{Liu14} Various theoretical proposals now exist for topological phases in (111) grown transition metal oxide films,\cite{Ruegg11_2,Yang:prb11a,Ruegg:prb12,Ruegg_Top:prb13,Hu:prb12,Okamoto:prb14,Yang:prl14,Okamoto:prl13,Xiao:nc11,Doennig:prb14,Lado:prb13,Liang:njp13,Wang:prb11,Wang14} and the list of candidate materials continues to grow.  We believe it is likely that experiment will indeed find evidence of topological phases in this class of systems.  Once a single example is found, experiments can be done to optimize material choices by ``perturbing" around this material with different isovalent elements, substrates, etc, in order to achieve the maximum bulk gap.  Our calculations suggest that the Chern insulator phase stands out as the mostly likely topological candidate for realistic conditions in thin film oxide heterostructures.

To be concrete, we will focus on the LaNiO$_3$ perovskite\cite{Ruegg11_2,Yang:prb11a,Ruegg:prb12,Ruegg_Top:prb13} and the Y$_2$Ir$_2$O$_7$ pyrchlore iridate.\cite{Hu:prb12,Yang:prl14}  (We note that an interesting theoretical study of Co-doped LaNiO$_3$ (111) grown bilayers suggest that correlation-driven odd parity superconductivity may appear in this system.\cite{Bing:prb14})
First principles calculations show that the bands around the Fermi energy are predominately of $d$-orbital character in both LaNiO$_3$ and Y$_2$Ir$_2$O$_7$.\cite{Ruegg:prb12,Ruegg_Top:prb13,Wan:prb11}  As a result, the spatial shape of the orbitals are highly asymmetric and can lead to interesting band features on the triangular, kagome, and honeycomb lattices that appear in the thin film structures of interest to us.  In particular, a simple tight-binding model for the Ni $d$-orbitals leads to flat bands that touch dispersing bands quadratically\cite{Ruegg11_2} (the flat touching feature persists with more realistic first principles band calculations\cite{Ruegg:prb12,Ruegg_Top:prb13}), and also bands that cross in a Dirac point (see Fig.\ref{fig:relaxed_BS}).  The stability of such band touching and crossing points with respect to interactions has been discussed recently in the literature in the context of interaction-generated topological phases.\cite{Raghu:prl08,Zhang:prb09,Sun:prl09,Wen:prb10,yu2010,Ruegg11_2,Yang:prb11a} 

 The central idea is that certain types of interactions, such as an on-site multi-orbital\cite{Ruegg11_2,Yang:prb11a} interaction or different site density-density interaction in a single band\cite{Raghu:prl08,Zhang:prb09,Sun:prl09,Wen:prb10} model, can generate an effective spin-orbit term (at the Hartree-Fock level) that favors topological phases.  This is one of the ideas we will explore in the remainder of this article in the context of real materials, though we will find that for heavier transition elements, such as iridium,  ``interaction generated spin-orbit coupling terms" are not needed to access topological phases.  Thus, any doubts about the reliability of the Hatree-Fock predictions for enhanced spin-orbit effects can be circumvented by focusing on classes of materials with intrinsically large spin-orbit coupling.

\subsection{LaNiO$_3$ bilayers}

\begin{figure}[htn]
\includegraphics[width=.9\linewidth]{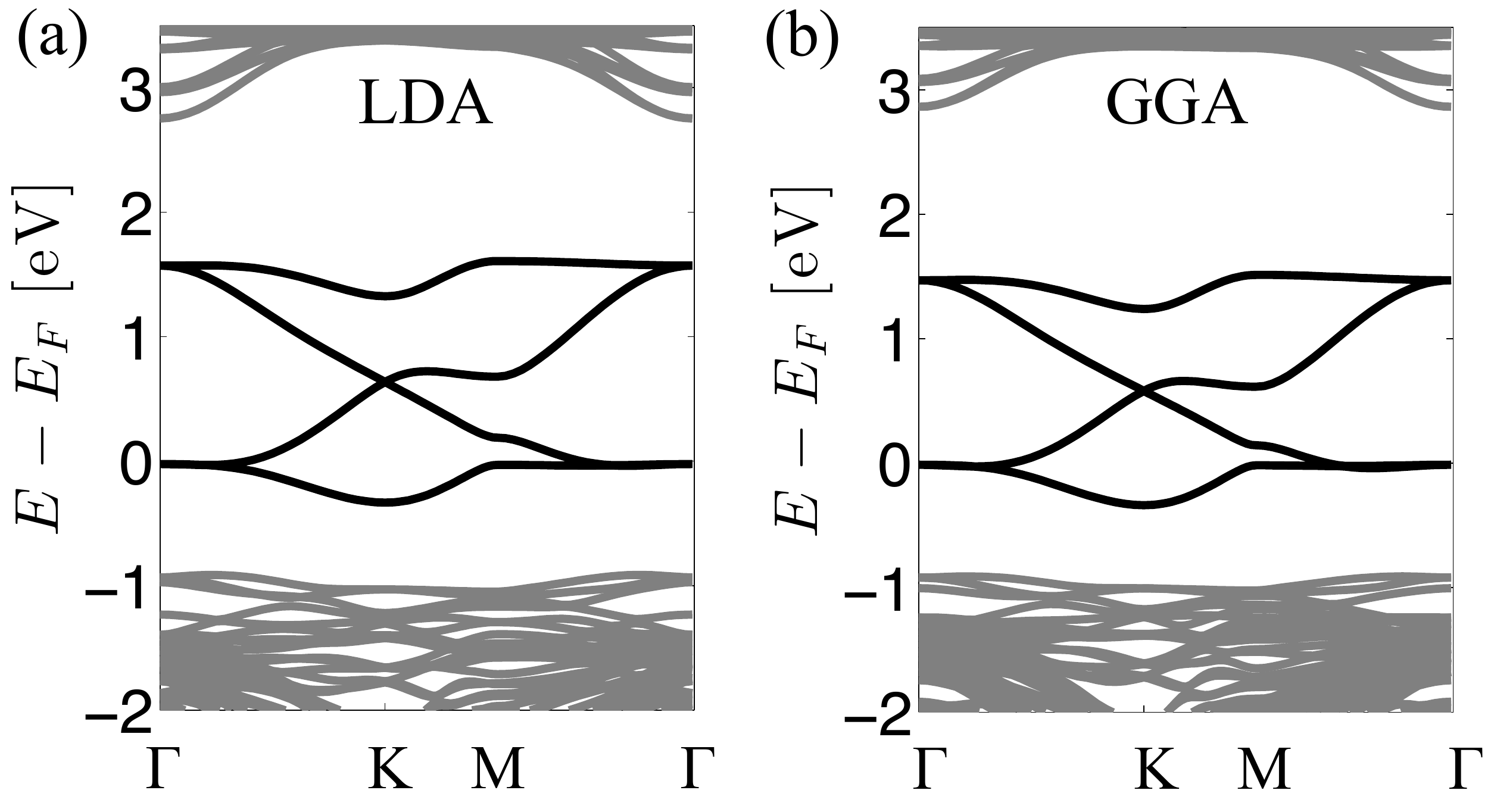}
\caption{First principles band structure of the fully lattice relaxed (LaAlO$_3$)$_{10}$/(LaNiO$_3$)$_2$/(LaAlO$_3$)$_{10}$ system within (a) LDA and (b) GGA, which are essentially indistinguishable. Note the quadratic band touching at the $\Gamma$ point as well as the linear crossings at K and K$'$.\cite{Ruegg_Top:prb13}  For a spin unpolarized system the Fermi energy lies right at the quadratic band touching point, while for a fully polarized (ferromagnetic) system the Fermi energy lies right at the Dirac point.  Our Hartree-Fock calculations suggest that for realistic interaction values, the system is very close to a fully spin-polarized state with a quantized Chern number--a quantum anomalous Hall state.\cite{Ruegg:prb12,Ruegg_Top:prb13}  See Fig.\ref{fig:relaxed_fit_phasediagram}(b).
}
\label{fig:relaxed_BS}
\end{figure}

\begin{figure}[t]
\centering
\subfigure[\ $\phi$ about (111)]{
\label{fig1a}
\includegraphics[width=0.25\linewidth]{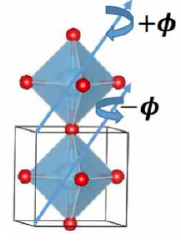}}
\subfigure[\ Title angle $\phi$ with layer index]{
\label{fig1b}
\includegraphics[width=0.65\linewidth]{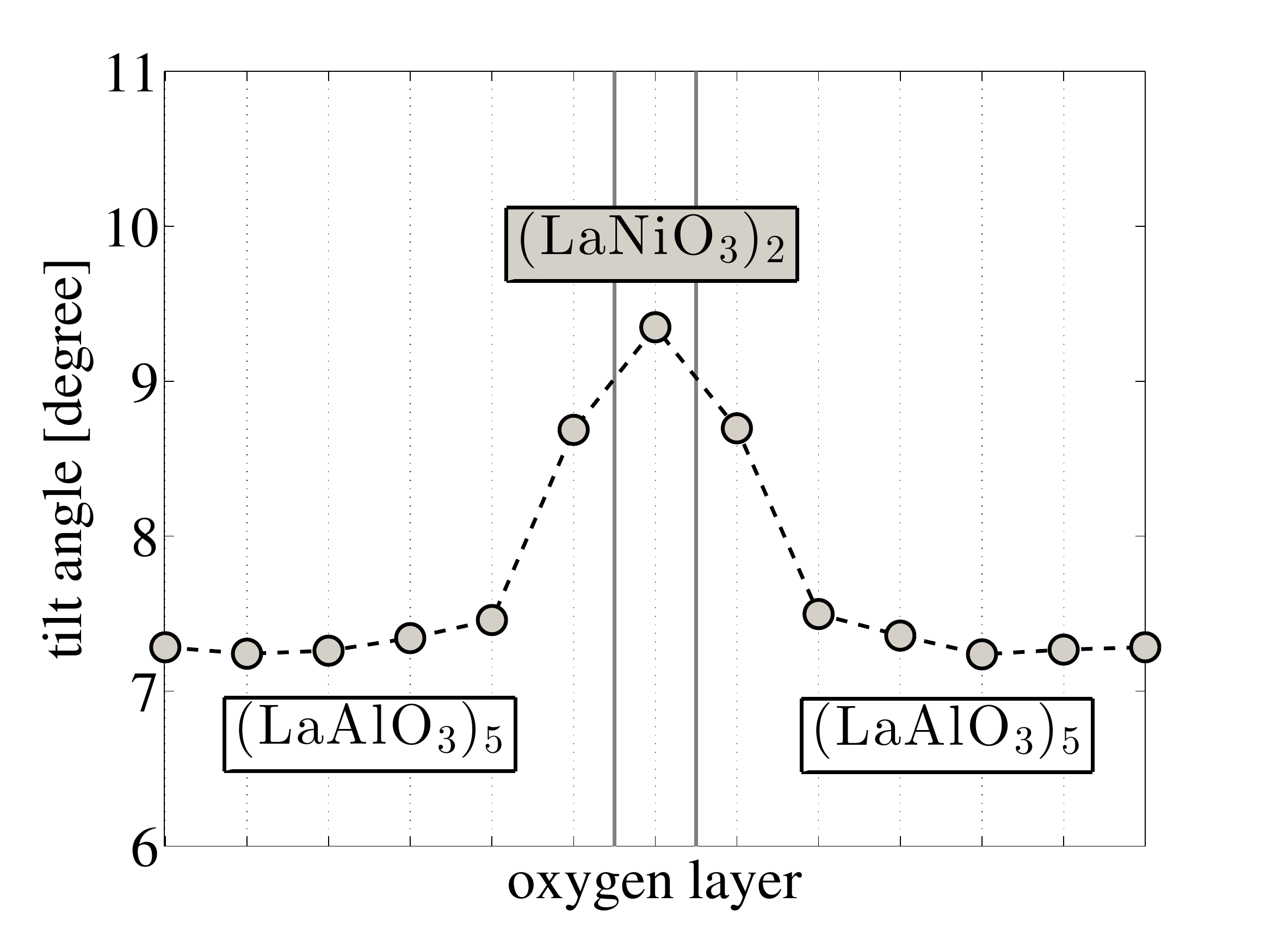}}
\caption{(Color online.)  (a) The pattern of the octahedral tilts/counter rotations present in the fully relaxed structure. (b) Layer resolved octahedral rotation angles for the (LaNiO$_3$)$_2$/(LaAlO$_3$)$_{10}$ supercell obtained within the LDA approximation to DFT.  As Fig.~\ref{fig:relaxed_BS} shows, these rotations do not lift the quadratic band touching at the $\Gamma$ point or the Dirac points at K, K' in the Brillouin zone because they preserve the trigonal point group symmetry.\cite{Ruegg_Top:prb13} This, in turn, implies the predictions for interaction-driven topological phases in the (LaNiO$_3$)$_2$/(LaAlO$_3$)$_N$ system remain qualitatively unchanged compared to the ``ideal" lattice structure.\cite{Ruegg:prb12}}
\label{fig:rotations}
\end{figure}

The band structure obtained with the local density approximation (LDA) and the generalized gradient approximation (GGA)\cite{Ruegg_Top:prb13} for the fully relaxed LaNiO$_3$ (111) bilayer is shown in Fig.~\ref{fig:relaxed_BS}.   The two are nearly indistinguishable.  Rotations of the octahedral oxygen cages are known to be important for large classes of transition metal oxides,\cite{TMO_book} including LaNiO$_3$ for which the adjacent oxygen cages counter-rotate about the (111) axis, as shown in Fig.\ref{fig:rotations}(a).  To perform Hartree-Fock calculations with this band structure,
we consider a tight-binding model based only on the nickel $e_g$ orbitals that includes nearest-neighbor hopping via the oxygen $p$-orbitals and second-neighbor hopping via the oxygen $p$-orbitals.  We find a a good fit by including the small differences in the hopping to ``outer" versus ``inner" oxygen atoms.\cite{Ruegg:prb12}  Assuming trigonal symmetry is preserved (a result consistent with our fully relaxed DFT results), we take the nearest-neighbor Slater-Koster parameters for hopping along the $z$-direction to be described by the matrix
\begin{equation}
\hat{t}_z=-\begin{pmatrix}
t&0\\
0&t_{\delta}
\end{pmatrix}
\label{eq:tz}
\end{equation}
in the basis $(d_{z^2},d_{x^2-y^2})$.  Here $t$ includes predominantly the hopping via the intermediate oxygen while $t_{\delta}$ arises from the direct overlap and is small. We set $t_{\delta}=0$ in the following. Assuming that the nearest-neighbor hopping in the $x$ and $y$ directions are equivalent to the hopping along the $z$ direction, we obtain the corresponding matrices by a rotation of the $e_g$-orbitals around (111) by $\pm 2\pi/3$. The matrix for the rotation by $2\pi/3$ is
$\hat{R}=\begin{pmatrix}
-1/2&\sqrt{3}/2\\
-\sqrt{3}/2&-1/2
\end{pmatrix}.$
As a result, we find $\hat{t}_x=\hat{R}^T\hat{t}_z\hat{R}, \quad \hat{t}_y=\hat{R}^T\hat{t}_x\hat{R}$. The Slater-Koster parameters for second-neighbor hopping define the matrix\cite{Ruegg:prb12}
\begin{equation}
\hat{t}_{xy}=-\begin{pmatrix}
t'/2&\sqrt{3}\Delta/2\\
-\sqrt{3}\Delta/2&-3t'/2
\end{pmatrix}.
\label{eq:txy}
\end{equation}
The off-diagonal entries proportional to $\Delta$ are allowed in the bilayer system discussed here (as opposed to a perfect cubic system) because the two possible paths connecting second-neighbor transition-metal ions are not equivalent: they either involve ``inner" or ``outer" oxygens.\cite{Ruegg:prb12} Note that $\hat t_{xy}$ is not symmetric if $\Delta\neq 0$ which means that there is an associated direction for the hopping. We use the convention that $\hat t_{xy}$ denotes the hopping of an electron along a second neighbor bond which is reached by first following the $y$-axis and then the $x$-axis of the cube.
By rotating the orbitals, we also obtain the second-neighbor hopping along the other directions:
$\hat{t}_{yz}=\hat{R}^T\hat{t}_{xy}\hat{R},\quad \hat{t}_{zx}=\hat{R}^T\hat{t}_{yz}\hat{R}.$

The generalized tight-binding model now takes the form
\begin{eqnarray}
H_0&=&\sum_{{\bs r}\in A}\sum_{s}\sum_{u=xyz}\left(\vec{d}^{\dag}_{s,{\bs r}}\hat{t}_u\vec{d}_{s,{\bs r}+{\bs e}_u}+{\rm h.c.}\right)\nonumber\\
&+&\sum_{{\bs r}\in A}\sum_{s}\sum_{u=xyz}\left(\vec{d}^{\dag}_{s,{\bs r}}\hat{t}_{u,u+1}\vec{d}_{s,{\bs r}+{\bs e}_u-{\bs e}_{u+1}}+{\rm h.c.}\right)\label{eq:H0}\\
&+&\sum_{{\bs r}\in B}\sum_{s}\sum_{u=xyz}\left(\vec{d}^{\dag}_{s,{\bs r}}\hat{t}_{u,u+1}\vec{d}_{s,{\bs r}-{\bs e}_u+{\bs e}_{u+1}}+{\rm h.c.}\right),\nonumber
\end{eqnarray}
where $\vec{d}_{s}=(d_{z^2,s},d_{x^2-y^2,s})^T$ is a vector in orbital space, $s=\uparrow$, $\downarrow$ is the spin and the notation $u+1$ refers to $y$ if $u=x$ with a cyclic extension to the other elements. 

\begin{table}
\begin{ruledtabular}
\begin{tabular}{l | l l l l l}
fit & $t$ [eV]& $t'$ [eV]& $\Delta$ [eV] & $E_F$ [eV]\\
\hline
unrelaxed (LDA) & 0.598 & 0.062 & -0.023 &  -0.693\\
fully relaxed (LDA)& 0.541 & 0.045 & -0.017 &  -0.641\\
fully relaxed (GGA)& 0.508 & 0.046 & -0.016 &  -0.593
\end{tabular}
\end{ruledtabular}
\caption{Parameters obtained in tight-binding fits to the $e_g$ DFT band structure of the unrelaxed and fully relaxed superlattice. There are very little changes between the unrelaxed and relaxed parameters, with $t'/t$ nearly invariant implying negligible change in the phase diagram shown in Fig.\ref{fig:relaxed_fit_phasediagram}.}
\label{tab:parameters}
\end{table}

\begin{figure}[htb]
\subfigure[\ LDA Tight Binding Fit]{
\includegraphics[width=0.45\linewidth]{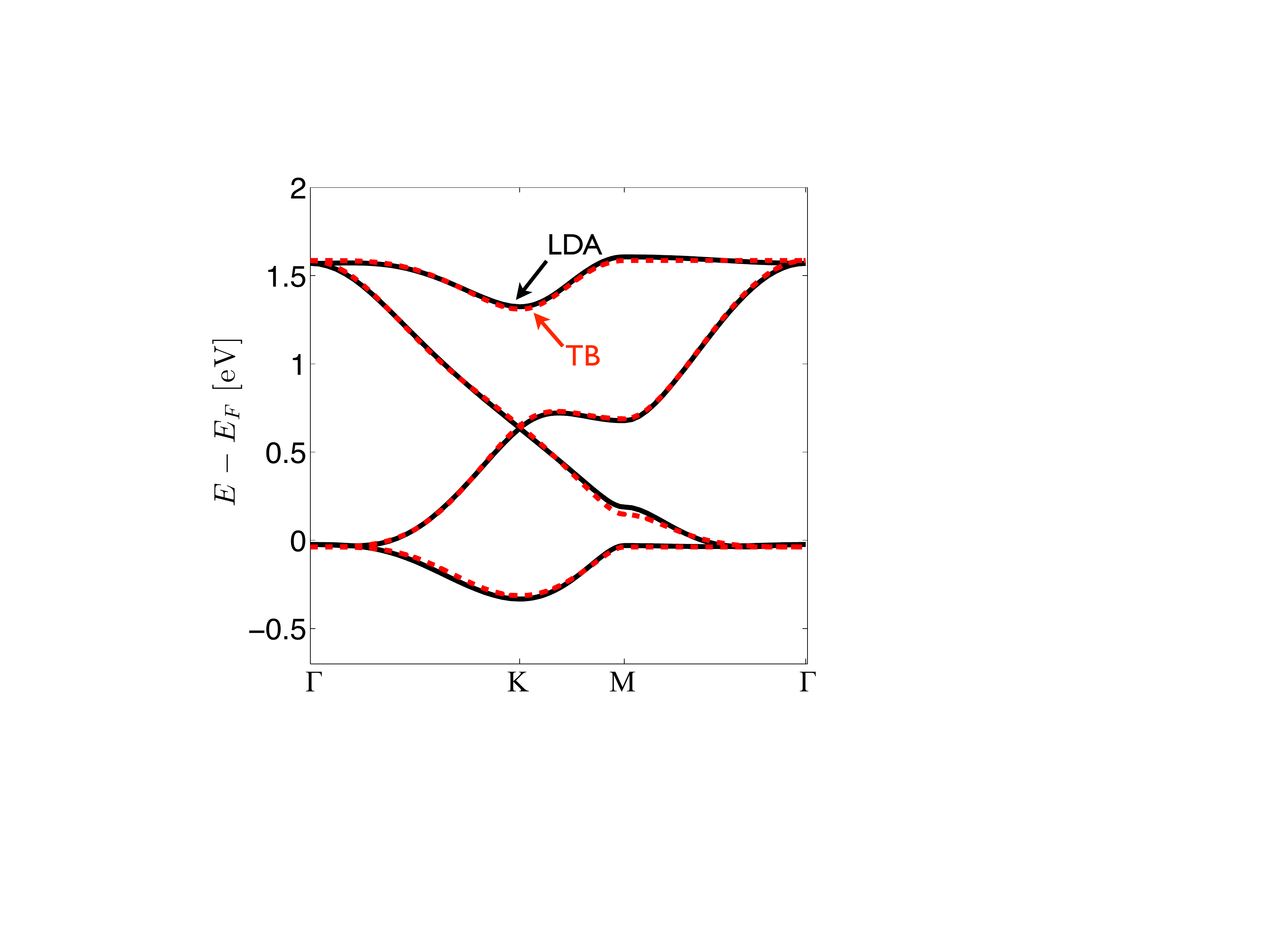}}
\subfigure[\ Hartree-Fock Phase Diagram]{
\includegraphics[width=0.47\linewidth]{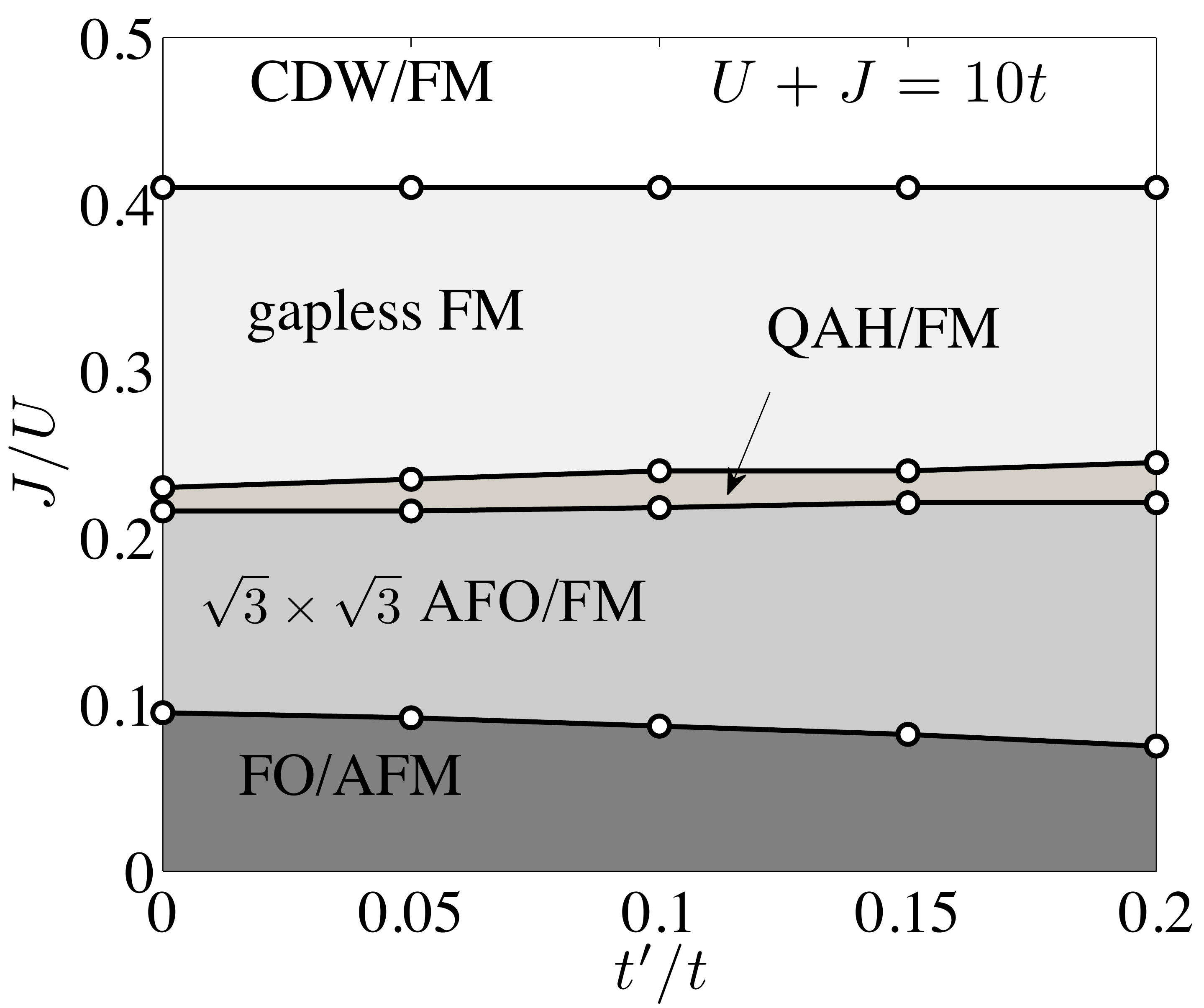}}
\caption{(color online) (a) Fully relaxed LDA band structure and tight-binding (TB) fit.  (b) Hartree-Fock phase diagram of the LaNiO$_3$ bilayer.  We estimate the experimental system has parameters $t'/t \approx 0.1$ and $J/U \approx 0.1-0.2$, which is rather close to the quantum anomalous Hall (QAH) phase with ferromagnetic order. FO=ferro-orbital, AFM=antiferromagnetic, AFO=antiferro-orbital, FM=ferromagnetic, CDW=charge density wave.
}
\label{fig:relaxed_fit_phasediagram}
\end{figure}

\begin{figure}[h]
\centering
\subfigure[\ Pyrochlore structure]{
\includegraphics[width=0.45\linewidth]{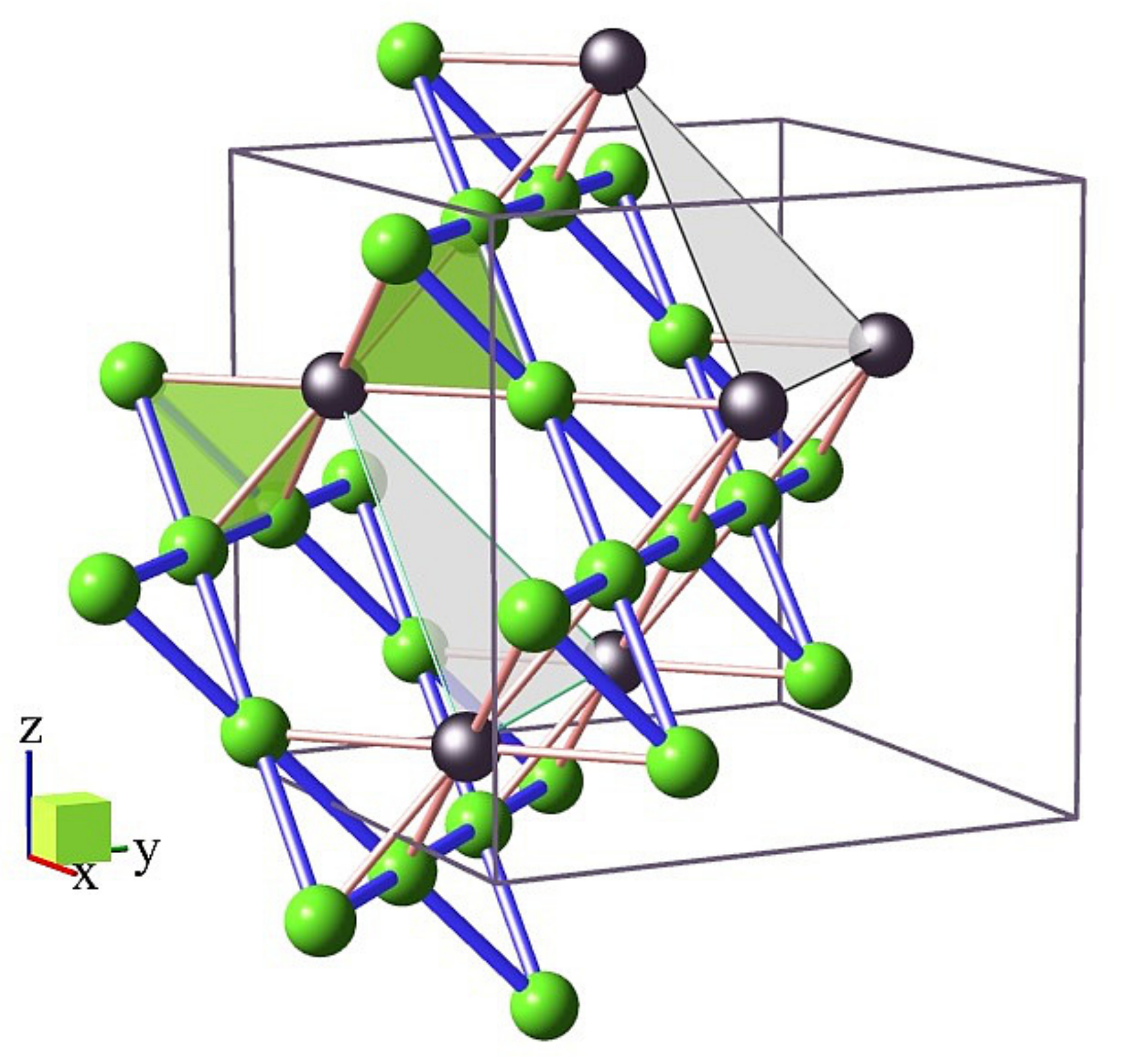}}
\subfigure[\ Bilayer structure]{
\includegraphics[width=0.45\linewidth]{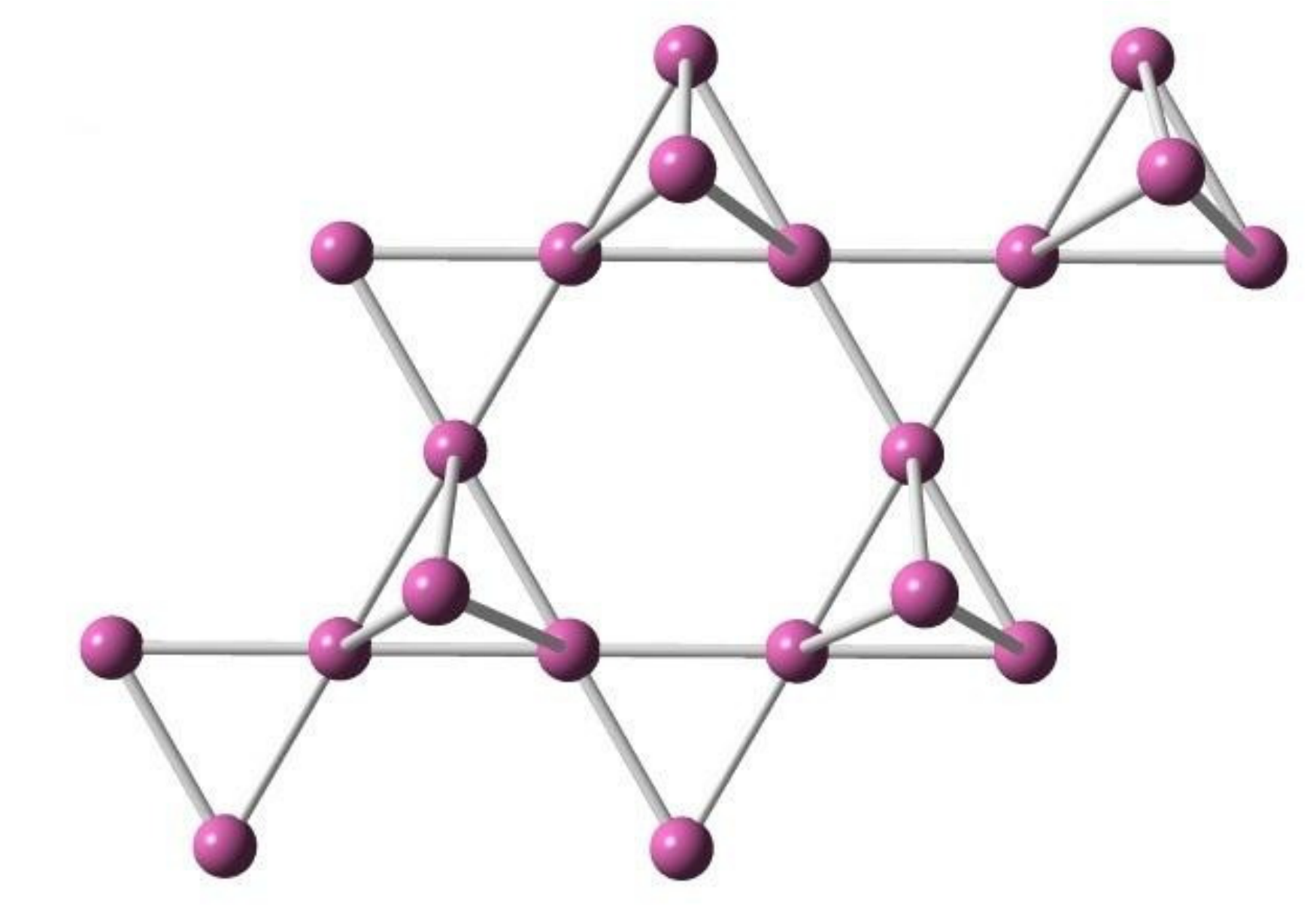}}
\caption{(Color online.)  (a) Bulk pyrochlore lattice structure showing the alternation of kagome planes (green balls on lattice sites) and triangular lattice planes (grey balls on lattice sites) along the (111) direction. (b) A bilayer film viewed from the (111) direction. We focus on the bilayer and the triangular-kagome-triangular structure, which show the most promise for topological phases. }
\label{fig:pyrochlore}
\end{figure}

Using the tight-binding model $H_0$ with parameters $t$, $t'$, and $\Delta$ (with $t_\delta=0$), we fitted both the LDA and GGA band structures of the fully relaxed system near the Fermi level. The fitting parameters are listed in Table~\ref{tab:parameters} and Fig.~\ref{fig:relaxed_fit_phasediagram}(a) shows the LDA together with the tight-binding band structure for the best fit.  To study the multi-orbital interactions within Hartree-Fock theory, we use an on-site interaction of the form\cite{Mizokawa:prb96,Imada:rmp98}
\begin{eqnarray}
H_{\rm int}&&=\sum_{\bs r}\Big[U\sum_ {\alpha}n_{{\bs r}\alpha\uparrow}n_{{\bs r}\alpha\downarrow}+(U'-J)\sum_{\alpha>\beta,s}n_{{\bs r}\alpha s}n_{{\bs r}\beta s}\nonumber\\
&&+U'\sum_{\alpha\neq \beta}n_{{\bs r}\alpha\uparrow}n_{{\bs r}\beta\downarrow}+J\sum_{\alpha\neq \beta}d_{{\bs r}\alpha\uparrow}^{\dag}d_{{\bs r}\beta\uparrow}d_{{\bs r}\beta\downarrow}^{\dag}d_{{\bs r}\alpha\downarrow}\nonumber\\
&&+I\sum_{\alpha\neq \beta}d_{{\bs r}\alpha\uparrow}^{\dag}d_{{\bs r}\beta\uparrow}d_{{\bs r}\alpha\downarrow}^{\dag}d_{{\bs r}\beta\downarrow}\Big].
\label{eq:Hint}
\end{eqnarray}
We assume $U'=U-2J$ and $I=J$, which are valid in free space and believed to be approximately true in the solid state environment. The total multi-orbital Hubbard Hamiltonian for the $e_g$ electrons is given by $H=H_0+H_{\rm int}$,
where $H_0$ is  given in Eq.~\eqref{eq:H0}.  The results\cite{Ruegg:prb12,Ruegg_Top:prb13} are shown in Fig.\ref{fig:relaxed_fit_phasediagram}(b). Studying a model that explicitly includes oxygen $p$-orbitals (where charge-transfer physics may appear) leads to similar results.\cite{Ruegg:prb12}

\subsection{Y$_2$Ir$_2$O$_7$ bilayers and trilayers}

In order to study the (111) films of the pyrochlore in a set-up similar to that shown in Fig.~\ref{fig:fig1}(a), we consider a tight-binding model
\begin{equation}
\label{eq:H0_p}
H_0=\sum_{\langle i,j \rangle,\alpha,\beta}t_{i\alpha, j\beta}c_{i\alpha}^{\dagger}c_{j\beta}-
\lambda\sum_i {\bf l}_i \cdot {\bf s}_i,
\end{equation}
where the $d$-orbital hopping takes the form\cite{Go:prl12,Witczak:prb12} 
\begin{equation}
\label{eq:t}
t_{i\alpha, j\beta}=t_{i\alpha, j\beta}^{in}+t_{i\alpha, j\beta}^{dir},
\end{equation}
which contains both an indirect and a direct hopping term between the $d$-orbitals.\cite{Pesin:np10,Yang_Kim:prb10,Kargarian:prb11} Here, $\lambda>0$ is the intrinsic spin-orbit coupling in the system which acts within the $t_{2g}$ manifold so $|{\bf l}|=1$, and ${\bf s}_i$ is the spin of the electron in a $t_{2g}$ $d$-orbital on site $i$.\cite{Pesin:np10,Yang_Kim:prb10,Kargarian:prb11} In the 5$d$ oxides, the strength of the spin-orbit coupling is estimated to be 0.2-0.7~eV and the hopping strength is on the order of 0.4-0.6~eV.\cite{Shitade:prl09,Kargarian:prb11}  The hopping amplitude in Eq.~\eqref{eq:t} contains a direct $d$-$d$ hopping, $t_{i\alpha, j\beta}^{dir}$, in addition to the indirect hopping via the oxygen orbitals.\cite{Go:prl12,Witczak:prb12} The direct hopping is parameterized by the strength of the $\sigma$-bonds, $t_s$, and the $\pi$-bonds, $t_p$.  Following Refs.~[\onlinecite{Go:prl12}] and [\onlinecite{Witczak:prb12}], we consider a set of representative ratios to explore a realistic parameter space: We set  $t_p=-2t_s/3$ and consider the cases of $t_s=-t$ and $t_s=t$.  Our preliminary GGA calculations for the thin-films suggest the band structure is most similar to that for $t_s=-t$, though the further neighbor hopping is considerably more important for the 5$d$ orbitals in Y$_2$Ir$_2$O$_7$ than it is for the 3$d$ orbitals in LaNiO$_3$.  

\begin{figure}[th]
\includegraphics[width=.9\linewidth]{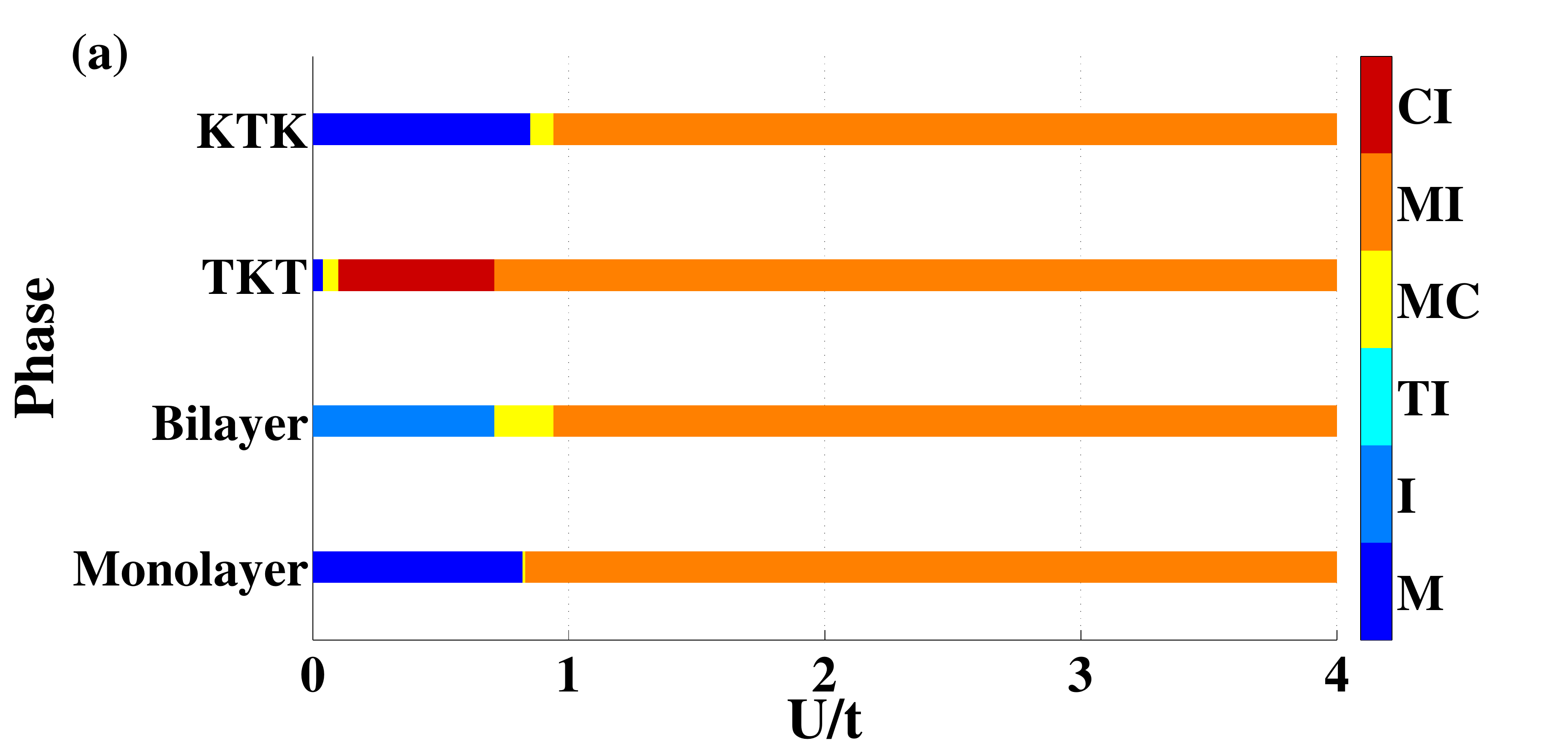}
\caption{(Color online.)  Hartree-Fock phase diagrams for the single kagome layer (monolayer), bilayer, triangular-kagome-triangular (TKT), and kagome-triangular-kagome (KTK) systems.  We have used  $t_s=-t$ and restricted ourselves to the $j=1/2$ manifold.  M=metallic, I=trivial insulator, TI=topological insulator, MC=magnetic conductor, MI=magnetic insulator, CI=Chern insulator (same as QAH).
}
\label{fig:pyro_phasediagrams}
\end{figure}

To carry out Hartree-Fock calculations for the Y$_2$Ir$_2$O$_7$ films, we use the tight-binding model in Eq.\eqref{eq:H0_p} with $t_s=-t$ supplemented by an on-site Hubbard term, $H_U=U\sum_i n_{i\uparrow}n_{i\downarrow}$, and further restrict ourselves to the $j=1/2$ manifold.\cite{Hu:prb12}  The results are shown in Fig.~\ref{fig:pyro_phasediagrams}.  Note the triangular-kagome-triangular (TKT) film supports a fairly wide region of a quantum anomalous Hall (Chern insulator=CI) state.  We find that if one includes fluctuations beyond the Hartree-Fock approximation, this region moves to larger $U$ values, close to what is reasonable for Y$_2$Ir$_2$O$_7$.  Therefore, we conclude that both the (111) grown LaNiO$_3$ bilayer and the (111) grown Y$_2$Ir$_2$O$_7$ TKT trilayer are candidates for a quantum anomalous Hall state.  Moreover, we find that small changes to the kinetic terms in Eq.\eqref{eq:H0} can lead to a $Z_2$ topological insulator in the (111) grown Y$_2$Ir$_2$O$_7$ bilayer for small $U$.\cite{Hu:prb12}

\section{Conclusions}

In conclusion, we have shown both the (111) grown LaNiO$_3$ bilayer and the (111) grown Y$_2$Ir$_2$O$_7$ TKT trilayer are candidates for a quantum anomalous Hall state, which will show a quantize Hall conductance.  We also found that realistic changes to the kinetic terms of Y$_2$Ir$_2$O$_7$ can also lead to a $Z_2$ topological insulator in the (111) grown Y$_2$Ir$_2$O$_7$ bilayer.  The most natural way to detect these states is through transport measurements, as has been done for the few known $Z_2$ TI\cite{Konig:sci07,Roth:sci09} and QAH systems.\cite{Chang:sci13}  From the point-of-view of technology applications, there is clearly a need in the field for more systems known to possess these phases.  The huge variety of transition metal oxides, combined with existing theoretical results, would seem to suggest it may only be a matter of time before a topological phase is discovered in a system similar to those we considered here.

\acknowledgements
We are grateful to our collaborators in this area, A.A. Demkov,  X. Hu, P. Jaduan, M. Kargarian, C. Mitra, and Z. Zhong.  Our work was generously funded by ARO Grant No. W911NF-09-10527, NSF Grant No. DMR-0955778, and DARPA grant No. D13AP00052. The Texas Advanced Computng Center (TACC) at The University of Texas at Austin for provided the necessary computing resources. URL:http://www.tacc.utexas.edu.

%

\end{document}